\documentclass{article}
\usepackage[utf8]{inputenc}

\usepackage{lipsum}
\usepackage{latexsym}
\usepackage{subfigure}
\usepackage[algo2e,ruled,vlined]{algorithm2e} 
\usepackage{amsmath}
\usepackage{algpseudocode}
\usepackage{algorithm}
\usepackage[table]{xcolor}
\usepackage{aaai21}  
\usepackage{times}  
\usepackage{helvet} 
\usepackage{courier}  
\usepackage[hyphens]{url}  
\usepackage{graphicx} 
\usepackage{natbib}  
\usepackage{caption} 
\title{SocialVec: Social Entity Embeddings}
\author{Nir Lotan, Einat Minkov }
\affiliations {
    \\
    University of Haifa \\
    nir.lotan@gmail.com, einatm@is.haifa.ac.il
}
\date{August 2020}

\begin{document}


\maketitle

\begin{abstract}
This paper introduces SocialVec, a general framework for eliciting social world knowledge from social networks, and applies this framework to Twitter. SocialVec learns low-dimensional embeddings of popular accounts, which represent {\it entities} of general interest, based on their co-occurrences patterns within the accounts followed by individual users, thus modeling entity similarity in socio-demographic terms. Similar to word embeddings, which facilitate tasks that involve text processing, we expect social entity embeddings to benefit tasks of social flavor. We have learned social embeddings for roughly 200,000 popular accounts from a sample of the Twitter network that includes more than 1.3 million users and the accounts that they follow, and evaluate the resulting embeddings on two different tasks. The first task involves the automatic inference of personal traits of users from their social media profiles. In another study, we exploit SocialVec embeddings for gauging the political bias of news sources in Twitter. In both cases, we prove SocialVec embeddings to be advantageous compared with existing entity embedding schemes. We will make the SocialVec entity embeddings publicly available to support further exploration of social world knowledge as reflected in Twitter.
\end{abstract}

\section{Introduction}
World knowledge about entities and the relationships between them is vital for information processing and communication by humans and machines alike. Much effort has been invested over the last decades in constructing factual knowledge bases that describe entities and the relationships between them in a structured relational form, e.g.,~\cite{mitchell2018never}. Following the advances in deep learning, researchers have proposed schemes for learning entity and relationship embeddings, based on the information available in structured knowledge bases~\cite{pbg19}, as well based on the textual contexts that surround the entity mentions~\cite{yamadaEMNLP2020}. 

However, much of the knowledge that is needed for intelligent information processing and communication extends beyond relational facts, and is in fact of social nature. Consider for example the social aspect of political polarity. Relevant knowledge about the stand of individuals who dissiminate information, or with whom one engages in a conversation, is necessary for effective information processing and communication. Likewise, the political leaning  of news sources should be taken into account for critical processing of the content disseminated by them~\cite{anICWSM12}.

In this work, we introduce {\it SocialVec}, a general framework for eliciting world knowledge from social networks. SocialVec learns entity representations based on the social contexts in which they occur within the social network. We apply this framework to Twitter,\footnote{https://twitter.com/} a popular and public social networking service that is considered as a credible source of social information (e.g., ~\cite{oconnorICWSM10}). We rely on the fact that public entities, including {\it politicians}, {\it artists}, national and local {\it businesses}, and so forth, maintain active presence in social networks in general, and Twitter in particular~\cite{marwick2011see}. We further exploit the fact that Twitter users typically {\it follow} other accounts of interest, to consume the content posted by those accounts.  We focus our attention on the most popular accounts within a large sample of the social network, assuming that these accounts represent {\it entities} of general interest. Importantly, we consider entities that are co-followed by individual users to be contextually related in that they reflect the interests, opinions, and socio-demographics of each user; for example, users would typically follow entities of similar political orientation to their own~\cite{eady2019}. The proposed SocialVec framework employs neural computing to process this information into low-dimensional entity embedding representations. Similar to  Word2Vec~\cite{Mikolov2013}, which learns low-dimensional word representations from the words with which they cooccur in large text corpora, SocialVec learns entity representations based on other entities that users tend to co-follow, as observed in a large sample of the Twitter network. While word embeddings facilitate tasks that involve text processing, we expect social entity embeddings to benefit information processing tasks of social flavor, such as the exploration of entity similarity in the social space, providing useful representations for downstream applications, and potentially supporting researchers and practitioners in deriving various social insights.

In this work, we apply SocialVec to a large portion of the Twitter network, which includes more than 1.3 million users sampled uniformly at random, and the accounts that they follow. We then exploit and evaluate the entity embeddings produced by SocialVec in two empirical case studies. The first study applies to the task of automatically inferring the personal traits of users from their social media profiles. We show that modeling users in terms of the popular accounts followed by them yields state-of-the-art or competitive performance in predicting various personal traits, such as gender, race, and political leaning. In another case study, we exploit SocialVec embeddings for gauging the political bias of news sources in Twitter. An evaluation against the results of polls conducted by Pew research~\cite{pew2014,pew2020} shows high accuracy of our approach. In both studies, we show clear advantage of SocialVec over existing entity embeddings, which rely on structured and textual information sources~\cite{pbg19,yamadaEMNLP2020}.

In summary, this paper presents several main contributions: (1) We outline SocialVec, a new framework for learning social vector representations of popular entities from social media, and (2) we apply and achieve high-performance results on the tasks of personal trait prediction and the identification of political bias of news sources. (3) We make the SocialVec entity embeddings publicly available, and believe that this has the potential of making a significant impact in exploring social world knowledge as reflected in Twitter.

\section{Related work}
\label{sec:related}


The success of Word2Vec~\cite{Mikolov2013} has inspired many related works, both within and beyond the textual domain. In the networks domain, the models of DeepWalk~\cite{Perozzi2014} and Node2Vec~\cite{groverKDD16} learn node embeddings by sampling node sequences via random walks in the graph, and predicting selected nodes based on the representations of the neighboring nodes in the sequence. Here, we do not aim to learn embeddings for all nodes (users) in the very large Twitter graph. Rather, we aim learn world knowledge in the form of entity embeddings, where we consider a bipartite graph that includes sampled users and the popular accounts that they follow within the Twitter network. We practically model two-hop node sequences, moving from a node that denotes a popular account to a user node who follows that account, and then to another popular account followed by the same user. In this manner, we avoid sampling paths from the very large graph of the whole of Twitter, efficiently modeling relevant information within a sub-graph of Twitter. Our approach is closely related to {\it Item2Vec}~\cite{barkan2016}, a model that learns item embeddings from a bipartite graph of user-item rating history for recommendation purposes. They too compute the embeddings of items given the ratings of individual users, considering other items known to be liked by the same users as relevant contexts. They found that the Item2Vec embeddings outperformed SVD in recommendation, especially when the rating matrix was sparse. In our work, we experiment with both models of Word2Vec, namely CBOW and skip-gram, whereas they only explore the latter. We also model full contextual information as opposed to sampling by shuffling, as detailed below.

\subsection{Entity embeddings}

Using SocialVec, we elicit social world knowledge from social media, learning the representations of popular accounts, which likely correspond to entities of general interest. Accordingly, we compare our learned embeddings with existing entity encoding schemes, all of which rely on factual sources. Below, we describe  and motivate the comparison of SocialVec embeddings with the the entity representations of Wikipedia2Vec and Wikidata graph embeddings.

\paragraph{Wikipedia2Vec}

Wikipedia is considered to be a high-quality semi-structured resource for learning entity representations.\footnote{https://www.wikipedia.org/}  In this space, entities correspond to concepts represented by dedicated Wikipedia articles. A useful feature of Wikipedia is the availability of human-curated annotations of entity mentions within its articles, and the mapping of the entity mentions onto their unique identifiers via hyperlinks, pointing to  relevant textual contexts of the entity mentions. The Wikipedia2vec model learns the embeddings of both words and entities from Wikipedia, with the aim of placing semantically similar words and entities close to each other in a joint vector space~\cite{yamadaACL16,yamadaEMNLP2020}. Concretely, this model learns word representations using Word2Vec, predicting the neighboring words of a given word in all of Wikipedia pages. In addition, for each hyperlink in Wikipedia, Wikipedia2Vec aims to predict the words that surround it given the referenced entity, thus modeling word-entity relationships, and further predicts the neighboring entities of each entity in Wikipedia’s link graph, i.e., the entities with which it is connected over some hyperlink, thus modeling direct inter-entity similarity. The resulting word and entity embeddings have been applied to various downstream tasks of natural language and knowledge processing, including entity linking, and knowledge base completion, and have been shown to outperform multiple baselines~\cite{yamadaEMNLP2020}.

\paragraph{Wikidata graph Embeddings}

Wikidata is a popular large collaborative knowledge base developed and operated by the Wikimedia Foundation~\cite{pellissierWWW16}, hence it is aligned with Wikipedia.\footnote{https://www.wikidata.org/} Wikidata represents entities of various types as nodes, and relational facts as typed edges between entity pairs. Additional relation types encoded in Wikidata include taxonomic hierarchies ('is a' relationships), and mappings to entity properties. Recently, Lerer {\it et al}~\citeyearpar{pbg19} 
have introduced the scalable PyTorch-BigGraph (PBG) framework, designed to efficiently learn graph-based node embeddings in very large graphs. We consider the entity embeddings inferred using PGB and the TransE~\cite{bordesNIPS13} graph embedding method from the whole of Wikidata. As reported by the authors, their implementation of TransE to Wikidata yielded higher-quality embeddings compared with the DeepWalk algorithm, as evaluated on the task of link prediction.

Relying on curated knowledge sources like Wikipedia or Wikidata has limitations however. First, these resources are inherently incomplete, where some popular entities do not have Wikipedia pages~\cite{hoffartCIKM12}. Similarly, the modeling of entity relationships and properties by these resources is partial~\cite{mitchell2018never}. Here, we further argue and show that these methods lack the representation of social world knowledge about entities.  Finally, we note that while several researchers have previously learned user embeddings from social network structure, these efforts were typically ad-hoc, being applied to specific tasks and datasets of limited size. Some related works consider the content associated with users for learning user embeddings, e.g., ~\cite{bentonACL16}. To the best of our knowledge, this is the first work that outlines and evaluates an approach for learning entity embeddings from social media at scale with the aim of capturing social world knowledge.

\section{Learning social embeddings}

Users on social networks typically associate themselves with other accounts of interest, consuming the content posted by those accounts. These directed social links correspond to a graph structure, where vertices denote user accounts and edges represent follower-to-followee relationships. Naturally, a fraction of the accounts have a high number of followers, where there exists a long-tail of accounts that are followed by small social circles. We focus our attention on those accounts that are most popular, assuming that they represent public {\it entities} of general interest. Our goal is to learn low-dimensional entity representations  that characterise the social contexts in which the entities appear on social media. 

Concretely, we consider a large sample of users $U$ mapped to the set of Twitter accounts that each one of them follows, $u_i \rightarrow \{a_{ij}\}, u_i\in U$, $a_{ij} \in A$, where $A$ denotes the union of all the accounts that are followed by the users in our dataset. We define the set of entities of interest as the subset of the most popular accounts, $E \subset A$, for which there exist at least $k$ users who follow them in our dataset.

In modeling the social context of each entity $e\in E$, we focus our attention on the sets of other entities that are co-followed by the individual users in our sample, i.e., considering the sets $\{\{a_{ij}\}: e\in \{a_{ij}\}, a_{ij} \in E\}$. Overall, the set of entities followed by an individual user is expected to be small, representing their personal interests and tastes. We discard information about the identity of the users, merely treating them as coherent samples of social contexts. Similarly to text sequences, which contain words that are related to each other grammatically and topically, we expect the sets of entities followed by the individual users to form meaningful units of local social contexts. 

\subsubsection{Learning}
We follow closely the Word2Vec approach that learns word embeddings from context word information~\cite{Mikolov2013}, adapting it to learn social contexts of entities. Given unlabeled text corpora comprised of a sequence of words $(w_i)_{i=1}^K$, the skip-gram network variant of Word2Vec is trained to predict the neighboring words that surround each word $w_i$ in turn within a window of a fixed size $c$. The loss function of skip-gram is defined as:
\begin{equation}
L = -\sum_{i=1}^K \sum_{-c\leq j\leq c, c\neq 0} log P(w_j \mid w_i)
\label{eq:context}
\end{equation}
and the conditional probability $P(w_j\mid w_i)$ is defined using the following softmax function:

\begin{equation}
P(w_j\mid w_i) = {exp(u_i^T v_j) \over \sum_{k\in W} exp({u_i}^T v_{k}) }
\label{eq:exp}
\end{equation} 
where $u_i \in R^d$ and $v_i \in R^d$ are latent vectors that denote the target and context word representations in the vocabulary $w_i\in W$, respectively. 

Equation~\ref{eq:exp} is overly costly however, as it applies to the whole vocabulary $W$. To alleviate this cost, negative sampling replaces the softmax function with:

\begin{equation}
P(w_j\mid w_i) = \sigma (u_i^T v_j) \prod_{k=1}^N \sigma (-u_i^T v_k)
\label{eq:nsampling}
\end{equation} 
where $\sigma = {1 / {1+exp(-x)}}$, and $N$ is a parameter. Thus, it is aimed to distinguish the target word $w_i$ from a noise distribution that includes $N$ negative examples, typically sampled from the unigram distribution raised to the 3/4rd power~\cite{mikolovNIPS13}. Training is performed by minimizing the loss function using stochastic gradient descent. 

While Word2Vec captures the linguistic context of words, learning to predict the adjacent words in a sequence, we wish to model socio-demographic contexts of entities, predicting other popular accounts that are co-followed by individual users. Adapting the formulation of the skip-gram model, we replace the words $w_i,w_j$ with account pairs $e_i,e_j$, modifying Equation~\ref{eq:context} as follows: 
\begin{equation}
L = -\sum_{u_i\in U} \sum_{e_i,e_j\in \{e_{u_i}\}, e_i\neq e_j} log P(e_j \mid e_i)
\label{eq:mod}
\end{equation}
where $\{e_{u_i}\}$ denotes the set of entities followed by user $u_i$. 

Our formulation is similar to the model of Item2Vec~\cite{barkan2016}, which learns item embeddings from user-item rating data for recommendation purposes. In both cases, the context corresponds to a set as opposed to a sequence. They shuffle the context items multiple times in order to obtain diverse context representations within a local context size $c$ (Eq.~\ref{eq:context}). We apply non-stochastic modeling of the entity neighborhoods (Eq.~\ref{eq:mod}) by practically setting the window size to be large enough to include all of the co-followed entities for every focus entity $e_i$.  

In addition to the skip-gram neural model, we experiment with the CBOW network variant of Word2Vec. Using the CBOW formulation, the focus word $w_i$ is predicted given all of the neighboring words within the context window. Similarly, we learn entity embeddings by learning to predict each focus entity $e_i$ in turn from the representations of all of the co-followed entities per user, $\{e_{u_i}\}, e_i\neq e_j$.

\subsection{The Social Corpus}

It is desired to train the neural embeddings using data that is high-quality, representative, and abundant. We obtained a large number of unique Twitter user identifiers, sampled uniformly at random from a pool of users in the U.S. who posted tweets in the English language,\footnote{We sampled the user identifiers from a corpus of 600 million tweets posted authored by over 10 million users in 2015, that was acquired from Twitter for research purposes.} and retrieved the full list of accounts followed by each user using Twitter API.  Overall, we collected information about the accounts followed by a total of 1.3 (1.265) million distinct users. Our data, collected in the beginning of 2020, includes 1,236 million relationships, mapping the users to 90.4 million unique accounts that are followed by them.

\begin{figure}[t]
\centering
\includegraphics[width=1\columnwidth]{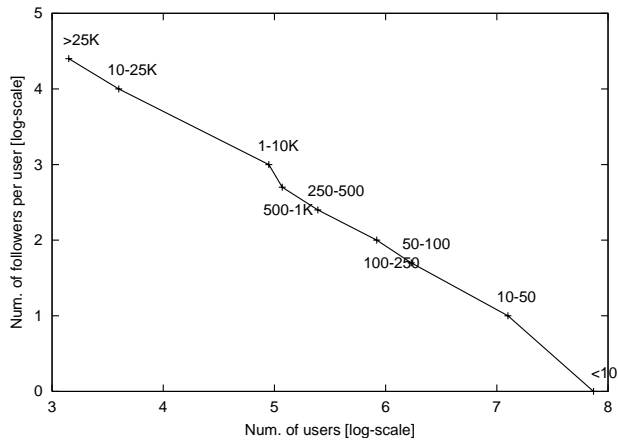}
\small
\label{fig:stats}
\caption{Account popularity based on our sample of 1.3 million Twitter users: a small number of highly popular accounts are followed by more than 25K users, where a long tail of accounts are followed by up to 100 users. We learn the embeddings of the most followed $\sim$200K accounts.}
\end{figure}

Figure~\ref{fig:stats} shows the distribution of accounts by their number of followers in our data. As shown, a small number of accounts (1.4K) are followed by more than 25K users, i.e., by more than 1.9\% of the users in our dataset. A higher number of accounts ($\sim$4,000) are followed by 10-25K users, i.e., by 0.8-1.9\% of the users. Next, there are many more accounts (89K) that are followed by 1-10K users, and a long tail of accounts that are followed by less than 1K users. 
Learning embeddings for all of the users in our dataset is infeasible, both computationally and statistically, since sufficient context information is required for learning meaningful representations. We therefore set a threshold based on account popularity, considering the accounts that are followed by at least $k=350$ users in our dataset. Roughly 200K accounts (201,247) meet this condition, and they comprise our vocabulary of entities $E$.

Naturally, we expect there to be differences in the scope of entities represented by a knowledge base like Wikipedia and Twitter. We aligned these resources, exploiting the fact that Wikidata links entity entries with Wikipedia as well as with their Twitter account information.\footnote{See Wikidata property: P2002, twitter user numeric id property P6552. We used Wikidata's SPARQL query service available at https://pypi.org/project/qwikidata/ to retrieve the Wikipedia identifier of each Twitter account represented in SocialVec.} We found that 31.5K out of the 200K accounts represented in SocialVec (15.8\%) have a Wikipedia page or Wikidata entry associated with them. Naturally, many of those entities that are represented in Wikipedia are widely-known and hence, popular in Twitter. 

Overall, the users in our sample follow 978 other accounts on average, among which 228 are considered popular and represented in SocialVec. About half of the accounts that one follows and are represented by SocialVec map also to Wikipedia. Thus, there is a substantial overlap between SocialVec vocabulary of entities and Wikipedia. Nevertheless, as one may expect, factual knowledge bases like Wikipedia are limited in their coverage of the world knowledge that is represented in social networks.

\subsection{Experimental setup}

We learned entity embeddings using the skip-gram and CBOW variants of Word2Vec as implemented in gensim~\cite{rehurek2010}. 
In order to capture the full context of co-followed accounts by every user, we set the window size to $c=1000$ (Eq.~\ref{eq:mod}). We discarded the records of users who follow more than 1000 accounts, assuming that these users are non-selective, and may not represent coherent contexts for our purposes.

We applied common parameter choices in training the models, setting the initial learning rate to 0.03, and having it decrease gradually to a minimum value of 7$\times 10^{-5}$. We set the number of negative examples to $N=20$, downsampling popular accounts by a factor of 1e-5. We have experimented and tuned these parameters based on cross-validation results using the training portion of our personal trait prediction dataset. We further experimented with learning embeddings of varying sizes, and set the embedding size to 100 dimensions--similar to Wikipedia2Vec entity embeddings--as this choice yielded good results in our cross-validation tuning experiments. The training of the models was conducted using an Intel Core i9-7920X CPU @ 2.90GHz computer with 24 CPUs, 128GB RAM and an NVIDIA GV100 GPU.
Training the CBOW and skip-gram models lasted about five days and two weeks, respectively.

\subsubsection{Evaluation}
\label{sec:comaprison}

We evaluate SocialVec embeddings on two case studies, in which we use these embeddings as features in learning, as well as employ vector semantics to assess semantic similarity between entities.
In our experiments, we compare SocialVec embeddings with the Wikipedia2Vec entity representations learned from Wikipedia, and the graph-based entity embeddings learned from Wikidata; both methods have been shown to yield SOTA results on downstream tasks, as discussed above. Concretely, we experiment with a version of Wikipedia2Vec embeddings which we trained on a dump of Wikipedia in English from October 2020.\footnote{We used the code provided by Yamada {\it et al.}~\citeyearpar{yamadaEMNLP2020}.} The Wikidata graph is very large, containing 78M entities and $\sim$4K relation types, posing a computational challenge. We use pre-trained entity embeddings learned using the TransE method from the official dump of Wikidata as of 2019-03-06 with the scalable PyTorch-BigGraph framework~\cite{pbg19}.\footnote{\url{https://github.com/facebookresearch/PyTorch-BigGraph}} The Wikipedia2Vec and graph embeddings are of 100 and 200 dimensions, respectively.

\section{Personal Trait Prediction}

Researchers have long been exploring methods for automatically inferring the socio-demographic attributes of users from their digital footprints, mainly based on the content posted or consumed by them~\cite{youyouPNAS15,volkovaACL16}. Such personal information about users is beneficial for applications like personal recommendation~\cite{wassermanIUI17}, as well as for social analytics~\cite{mueller2021demographic}. Here, we employ the SocialVec embeddings of popular accounts that users follow on Twitter as features in a supervised classification framework, targeting the prediction of various socio-demographic traits for each user. 

\subsection{Dataset}

\begin{table}[t]
\small
\centering
\begin{tabular}{lll}
\hline
Attribute & Class Distribution & Profiles \\
\hline
Age & $\leq 25$ y.o. (56\%), $>25$ y.o. & 3,485 \\
Children & No (82\%), Yes & 3,485 \\
Education & High School (67\%), Degree & 3,485 \\
Ethnicity & Caucasian (57\%), Afr. Amer. & 2,905 \\
Gender & Female (56\%), Male & 3,475 \\
Income & $\leq$ 35K (64\%), $>$35K & 3,485 \\
Political & Democrat (76\%), Republican & 1,790 \\
\hline
\end{tabular}
\caption{Personal trait prediction: dataset statistics}
\label{tab:volk_stats}
\end{table}

We refer to a labeled dataset due to Volkova {\it et al.}~\citeyearpar{volkovaAAAI15}. This dataset includes the identifiers of sampled Twitter users, labeled by means of crowd sourcing with respect to the personal attributes of {\it age}, {\it gender}, {\it ethnicity}, {\it family status}, {\it education}, {\it income} level, and {\it political orientation}. The labels were determined based on public information on Twitter, including the user's self-authored account description, and any metadata and historical tweets available for the user. Thus, the labels reflect subjective and proximate judgements. All of the labels are binary, where continuous attributes, namely {\it age} and {\it income}, were manually split into distinct binary ranges, as detailed in Table~\ref{tab:volk_stats}.  

In order to obtain information about the accounts that the users in the dataset follow, we queried Twitter API with the relevant user account identifiers. Overall, we tracked 3,558 active user profiles in Twitter (out of 5,000 users in the source dataset).  We further retrieved tweets posted by these users,\footnote{Due to legal restrictions, public datasets specify user ids, but do not contain the content posted by them.} considering text as alternative information source for attribute prediction. Overall, we collected up to 200 tweets posted by each user, similar to Volkova {\it et al.}~\citeyearpar{volkovaAAAI15}, obtaining 180 tweets per user, on average.

While this data was collected some time after the user labels were assigned, we believe that as the labels are categorical, and were obtained based on coarse human judgement, the labeling accuracy is not severely compromised. Furthermore, we evaluate the various methods using the same data and in the same conditions, where this forms a viable evaluation setup.


\subsection{Methods}

We perform supervised classification experiments, predicting the various attribute values for each user as independent binary classification tasks. For each target attribute, we randomly split the set of labeled examples into distinct train (80\%) and test (20\%) sets, maintaining similar class proportions across these sets. Once the models are trained, prediction performance is evaluated against the gold labels of the test examples.

In learning, we represent each user $u$ using the vector embeddings of the popular accounts that they follow, $\{e_u\}$. We follow the practice of averaging the bag-of-embedding vectors into a unified summary vector of the same dimension $\overline{e}_u$ ~\cite{shenACL18}. We then feed the averaged vector representation into a logistic regression classification network, in which the output layer consists of a single sigmoid unit. While we experimented also with multi-layer network architectures, we found this single-layer classifier to work best. 

Below, we report our results applying this framework using SocialVec and the existing entity embeddings schemes. As reference, we evaluate also content-based attribute classification using the tweets posted by each user as relevant evidence. In this case, the tweets authored by each user, $t_u$, are first converted into a 300-dimension text embedding vector using the pre-trained convolutional FastText neural model~\cite{joulinEACL17},\footnote{https://github.com/facebookresearch/fastText} which is often used for tweet processing~\cite{soseaEMNLP2020}. We average the FastText tweet embeddings~\cite{adiICLR17}, and feed the resulting user representation to the logistic regression network.

\subsection{Results}

\begin{figure}[t]
\centering
\includegraphics[width=0.9\columnwidth]{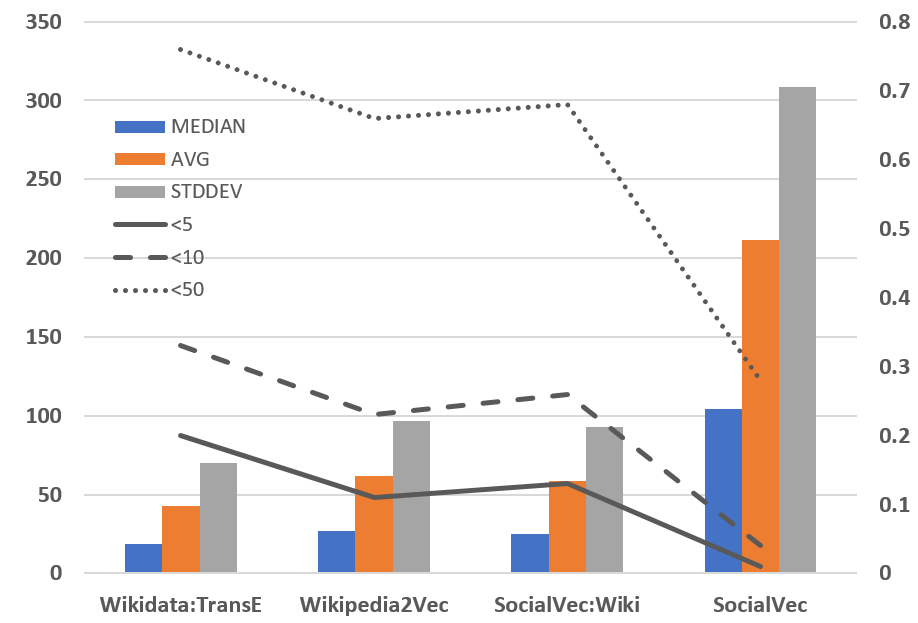}
\small
\caption{Personal trait prediction: the number of popular account embeddings that are associated with each user in the dataset, and the proportions of users that have a limited number of embeddings (less than 5,10, or 50) associated with them using each method.}
\label{fig:stats}
\end{figure}

\begin{table*}[t]
\small
\centering
\begin{tabular}{lllllllll}
\hline
 & Age & Children & Education & Ethnicity & Gender & Income & Political \\
\hline \hline
SocialVec & \textbf{0.738} & \textbf{0.683} & 0.739 & \textbf{0.953} & \textbf{0.890} & 0.732 & \textbf{0.798} \\
{\it Wikipedia entities:} & & & & & & & &  \\
Wikidata:TransE & 0.686 & 0.614 & 0.690 & 0.864 & 0.803 & 0.682 & 0.694 \\
Wikipedia2Vec & 0.614 & 0.610 & 0.628 & 0.704 & 0.641 & 0.635 & 0.599  \\
SocialVec $\cap$ Wiki. & 0.705 & 0.665 & 0.698 & 0.924 & 0.859 & 0.709 & 0.748 \\
{\it Content-based:} & & & & & & & &  \\
FastText & 0.695 & 0.575 & \textbf{0.740} & 0.785 & 0.768 & \textbf{0.748} & 0.654 \\
$^*$~\cite{volkovaACL16} & {\it 0.63} & {\it 0.72} & {\it 0.77} & {\it 0.93} & {\it 0.90} & {\it 0.73} & - \\
\hline
\end{tabular}
\caption{Personal trait prediction results [ROC AUC] ($^*$ the results by~\cite{volkovaACL16} were obtained on a larger version of our dataset, and are not directly comparable.)}
\label{tab:results}
\end{table*}

The classification results for each of the target attributes are given in Table~\ref{tab:results} in terms of the ROC AUC measure. Let us first examine our results using the SocialVec embeddings as features in classification, and compare these results with the Wikipedia2Vec and Wikidata entity embeddings. As detailed in the table, SocialVec outperforms the other embedding schemes by a large gap across all of the target attributes. For example, classification performance on age prediction is 0.74 in terms of ROC AUC using SocialVec, compared with 0.69 and 0.61 using the Wikidata graph embeddings and Wikipedia2Vec, respectively. Similar or larger gaps in favor of SocialVec are observed for each for the other traits.

Importantly, only a subset of the popular accounts that are represented by SocialVec have respective embeddings based on Wikipedia and Wikidata. Figure~\ref{fig:stats} shows relevant statistics about the number of accounts followed by the users in our dataset that are represented using each method. As illustrated, the number of relevant SocialVec embeddings that are available is substantially higher compared with the other methods. Consequently, the ratio of users that are poorly represented, being associated with less than 10 account embeddings, is 5\% using SocialVec, vs. 22 and 32\% using the Wikipedia- and Wikidata-based methods. These gaps in coverage illustrate the wider applicability of our approach in the social domain of Twitter. Yet, in order to conduct a fair comparison between the different embedding schemes, we performed another experiment, in which we eliminated from the feature any SocialVec embeddings of accounts that are not represented by either one of the other methods. As illustrated in Figure~\ref{fig:stats}, this variant has a similar coverage profile as the Wikipedia-based methods. 

The results of using the restricted set of SocialVec embeddings are shown in Table~\ref{tab:results} ('SocialVec $\cap$ Wiki.'). We observe that the classification performance using this strict evaluation is lower as less features are used, e.g., ROC AUC drops from 0.74 to 0.71 on the {\it age} category, however SocialVec still outperforms the performance of the alternative entity representations on each and every one of the target categories. Thus, we conclude that  our social entity embeddings inferred from Twitter are more informative compared with the respective entity embeddings inferred from relational knowledge bases for personal trait prediction.

\paragraph{Feature analysis.}

\begin{table*}[t]
\begin{footnotesize}
\centering
\begin{tabular}{l|l}
\hline
\textbf{Male} & \textbf{Female} \\
\hline
{\it Ian Rapoport}, Sports writer and analyst (1.04) & {\it Chelsea DeBoer}, a reality TV persona (0.81) \\
{\it Chris Broussard}, Sports analyst, Fox Sports (1.02) & {\it womenshumor}, "tweets made for a woman" (0.80) \\
{\it Adam Schefter}, Sports analyst (1.02) & {\it Maci Bookout}, a reality television personality (0.76) \\
\hline
\textbf{White} &  {\textbf{Afro-American}} \\
\hline
{\it starwars}, Star Wars on Twitter (0.80) & {\it KYLESISTER} (1.17) \\
{\it John Krasinski}, an actor, director and producer (0.78) & {\it Emmanuel Hudson}, actor (1.16) \\
{\it Luke Bryan}, a country music singer and songwriter (0.78) & {\it Erica Dixon}, TV personality (1.15) \\
\hline
\textbf{High-school} & \textbf{Academic} \\
\hline
{\it 21 Savage}, a rapper, songwriter, and producer (0.45) & {\it The New Yorker}, an American magazine (1.30) \\
{\it AccessJui}, Music Producer (0.44) & {\it The Economist}, an international newspaper (1.20) \\
{\it Desi Banks}, a comedian, actor, and writer (0.42) & {\it Jack Tapper}, anchor and host at CNN (1.19) \\
\hline
\textbf{Republican} & \textbf{Democratic} \\
\hline
{\it Chick-fil-A}, a large fast food restaurant chain (1.15) & {\it Bryson Tiller}, a rapper (0.35) \\
{\it Carrie Underwood}, a Country singer (1.14)  & {\it Kevin Gates}, a rapper (0.35) \\
{\it Tim Tebow} (1.13) & {\it Tami Roman}, a TV personality and rapper (0.34) \\
\hline
\end{tabular}
\caption{The top Twitter accounts that are characteristic to different subpopulations as measured using our datasets labeled with personal attributes and the Pointwise Mutual Information (PMI) measure. }
\label{tab:topAccounts}
\end{footnotesize}
\end{table*}

Table~\ref{tab:topAccounts} demonstrates in more detail the valuable social information that SocialVec encodes for personal trait prediction. The table presents the top accounts associated with each class label in our dataset, based on the pointwise mutual information (PMI) measure~\cite{rudingerETAL17}. In general, high PMI values indicate on distinctive feature-class correlation. We observe, for example, that the top Twitter accounts followed by male (as opposed to female) users belong to men who specialize in sports, whereas the top accounts that characterize female users belong to women. Likewise, we find that the top accounts that characterise Afro-American users belong to Afro-Americans, and vice versa. We further observe that users with an academic degree distinctively follow media accounts such as the New Yorker and the Economist magazines, whereas non-academic users tend to follow rappers.  Finally, distinctive accounts that represent political polarity include rappers on the Democratic side. On the Republican side, we find accounts that are related to Country music, echoing previous findings by which Country music fans are twice as likely to vote Republican than fans of other genres,\footnote{https://news.gallup.com/poll/13942/Music-Cars-2004-Election.aspx} and the accounts of Tim Tebow, a former football player, and the fast-food brand of Chick-fil-A, which are both known for their conservative views.\footnote{See "Tebowing" at https://en.wikipedia.org/wiki/Tim\_Tebow, and https://en.wikipedia.org/wiki/Chick-fil-A: Same-sex marriage controversy.} Thus, the accounts that one follows on social media are highly predictive of their personal traits and interests; this social information is encoded more effectively and with better coverage using SocialVec compared with the Wikipedia-based methods.

\paragraph{Comparison with other approaches}

A question of interest is how the modeling of users based on the entities that they follow as encoded by SocialVec compares with the more traditional approach of representing the users in terms of the content authored by them. Table~\ref{tab:results} presents also our trait prediction results based on the textual content authored by the users ('FastText'). The best results per trait are highlighted in boldface in the table. As shown, the SocialVec approach achieves top performance by a large margin on all categories, except for {\it education} and {\it income}, for which content-based classification achieves comparable or slightly higher results. (The difference in performance between FastText and SocialVec on the {\it education} category is not significant according to the McNemar ${\chi}^2$ statistical test as applied to accuracy results.) We find this sensible, as education and income levels are known to be manifested through writing style~\cite{flekovaACL16}. 

Finally, Table~\ref{tab:results} includes the results previously reported by Volkova {\it et al.}~\citeyearpar{volkovaAAAI15,volkovaACL16}. They trained log-linear models using n-gram features extracted from the tweets posted by each user, showing gains over alternative models. We stress that their results are not directly comparable to ours, as the original dataset included many more labeled examples of users who are no longer active (5K vs. 3.5K labeled users in our dataset). Moreover, they relied on tweets obtained shortly after user labeling. Nevertheless, we observe that our results applied to the reduced dataset using SocialVec are comparable or exceed the results by Volkova and Bachrach~\citeyearpar{volkovaACL16} on the majority of the categories. Another work that predicted {\it gender} and {\it ethnicity} from user names reported results that are lower than those reported by Volkova and Bachrach~\cite{wood18}.

In summary, our results indicate that many socio-demographic attributes can be predicted with high accuracy using SocialVec entity embeddings. We showed superior performance compared with entity embeddings inferred from knowledge bases, both due to better coverage and the encoding of social aspects in SocialVec. And, our results outperform content-based classification for multiple attributes, even when trained using fewer labeled examples. To improve prediction results further, it is possible to model the network information of  the users' direct friends, exploiting social homophily~\cite{panACL19}. Also, trait prediction results may potentially improve via the integration of network and content information, ideally using larger datasets. 

\section{Political polarity of news sources}

As a second case study, we investigate whether the political orientation of news sources can be inferred from the social patterns encoded in SocialVec. 
A recent survey by Pew Research estimated that 62\% of the U.S. adults consume news primarily from social media sites~\cite{mitchell2016key}. A lack of awareness of the biases of these accounts can play a critical role in how news are assimilated and spread on social media, shaping people’s opinions and influencing their choices to the extreme of swaying the outcomes of political elections~\cite{allcott2017}. In addition, identifying the slant of news accounts on social media may help  address political bubbles, where users are exposed primarily to ideologically congenial political information~\cite{eady2019}. 

Various research works aimed to infer the political slant of media sources based on the language used by them, the framing of political issues by these sources~\cite{balyACL20}, or the language used by their followers~\cite{stefanovACL2020}.\footnote{~\cite{stefanovACL2020} referred to somewhat disputable judgements by the mediaBiasFactCheck website. We could not obtain the subset of accounts that they evaluated for comparison purposes.} Ribeiro {\it et al.}~\citeyearpar{ribeiroICWSM18} quantified the biases of news outlets by analyzing their readership directly, considering the proportions of liberal and conservative users within the source’s audience. Their work is limited to Facebook, as they relied on explicit account statistics that it provides to advertisers. In this work, we exploit SocialVec social entity embeddings for predicting the political leaning of news accounts on Twitter. As we show empirically, the resulting assessments are highly accurate, yielding similar results to formal polls, whereas factual entity embeddings lack the necessary social information that is encoded by SocialVec.

\subsection{Methods}

The Word2Vec metric tends to place two words close to each other if they occur in similar contexts~\cite{levy2014neural}. Likewise, we gauge the social similarity between entities based on the cosine similarity of their embeddings. Assuming that individual users follow the accounts of politicians, media sources, and other entities with similar political orientation to their own, we expect the distribution of accounts that are followed by right- and left-leaning populations to be distinguishable. That is, the embeddings of entities of similar political orientation should exhibit higher similarity in the vector space, compared with the embeddings of accounts of opposite political polarity. We therefore compute the bias of news accounts on Twitter based on their similarity to popular accounts with distinct political polarity in the embedding space. Specifically, we consider the accounts of the Republican Donald Trump, the incumbent U.S. president at the time that our data was collected,\footnote{https://twitter.com/realDonaldTrump;  suspended in Jan. 2021.} and of Barack Obama, the former Democratic president.\footnote{https://twitter.com/BarackObama} As of 2020, both accounts were among the top-followed Twitter accounts in the U.S., ranked at fourth and first positions, respectively, based on the number of their followers.\footnote{https://www.socialbakers.com/statistics/twitter/profiles/united-states} 

Let us denote the SocialVec embedding of a specified news account as $e_n$, and the embeddings of the Democratic and Republican anchor accounts, which we set to the accounts of Obama and Trump, as $e_D$ and $e_R$, respectively. We measure the similarity of the news source in the embedding space with these Republican and Democratic anchors. We then assess the political orientation of the news account, considering the difference between those similarity scores. Formally, we compute the {\it political orientation} (PO) score of $e_n$ as the difference between the cosine similarities:
\begin{equation}
PO(e_n)=Sim(e_R,e_n)-Sim(e_D,e_n)
\label{eq:bias}
\end{equation} 
Accordingly, a positive score indicates on overall conservative (Republican)  social orientation, whereas a negative score indicates on a liberal (Democratic) social bias. The greater the gap between the similarity scores, the greater is the social political polarity. 

In our experiments, we rank selected news accounts according to their computed political orientation scores, and compare our results against formal polls. Again, we evaluate the embeddings of SocialVec, as well the embeddings of Wikipedia2Vec and the graph-based Wikidata embeddings, gauging the extent to which each of these methods captures the social phenomena of political leaning.

\subsection{Ground-truth datasets}

We refer to the results of two formal polls conducted by Pew Research in 2014 and 2019~\cite{pew2020}, with the goal of gauging the political polarization in the American public. The participants in the polls were recruited using random sampling of residential addresses, and the data was weighted to match the U.S. adult population by gender, race, ethnicity, education and other categories. In both polls, Pew researchers classified the audience of selected popular news media outlets based on a ten question survey covering a range of issues like homosexuality, immigration, economic policy, and the role of government. The media sources were then ranked, according to those poll participants who said they got political and election news there in the week before, taking into consideration the party identification (Republican or Democrat) and ideology (conservative, moderate or liberal) of those participants.

Overall, the polls conducted in 2014 and 2019 apply to 36 and 30 selected news media outlets, respectively. These two sets include 43 unique media outlets jointly, where there are 18 news sources that overlap between the two surveys. We manually mapped the various sources to their Twitter accounts, identifying the accounts for the majority of news sources included in the earlier poll (31 out of 36), and practically all of the news sources included in the more recent poll (30). As detailed in Table~\ref{tab:polls}, all of those Twitter accounts are included in SocialVec, where most of the news sources have respective  Wikipedia2Vec and Wikidata embeddings.

\subsection{Results}
 
The Pew surveys assign numerical political polarity score to each of the news sources that do not match the range, or interpretation, of the computed cosine similarity scores. In order to assess the correlation between our entity similarity metric and the survey results, we therefore consider the relative ranking of the various news sources, ranging from conservative/Republican to liberal/Democrat. 

\begin{figure*}[t]
\centering
\frame{\includegraphics[width=0.8\textwidth]{ 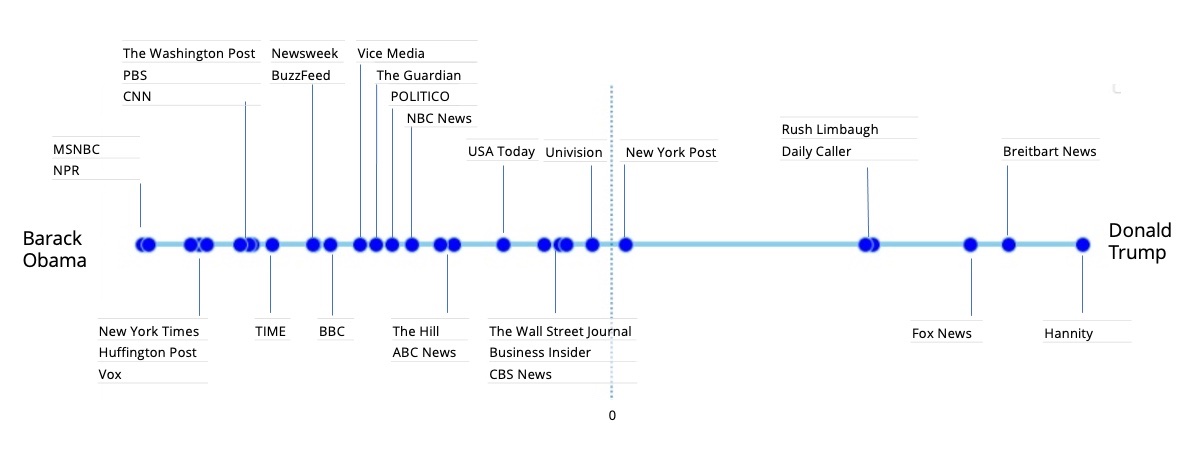}}
\caption{Ranking of political polarity based on our embeddings}
\label{fig:political}
\end{figure*}

\begin{table}[t]
\small
\centering
\begin{tabular}{lcccc}
\hline
Poll & \# Accounts & SocialVec & Wikipedia2Vec & Wikidata \\
\hline
2014 & 31 & \textbf{0.82} (31) & 0.36 (28) & -0.40 (23) \\
2020 & 30 & \textbf{0.85} (30) & 0.28 (28) & -0.32 (23) \\
\hline
\end{tabular}
\caption{Spearman's correlation results of ranking news accounts by political slant using different entity embeddings, compared with the poll-based rankings reported by Pew Research in 2014 and 2020. The number of available news account embeddings is given in parenthesis for each method.}
\label{tab:polls}
\end{table}

\begin{table}[t]
\small
\centering
\begin{tabular}{llcccc}
\hline
Poll & Accounts & SocialVec & Wikipedia2Vec & Wikidata \\
\hline
2014 & {\it All} & \textbf{0.94} (31) & 0.55 (28) & 0.32 (23) \\
& {\it Common} & \textbf{0.95} (22) & 0.55 (22) & 0.27 (22) \\
\hline
2020 & {\it All} & \textbf{0.97} (30) & 0.60 (28) & 0.27 (23) \\
& {\it Common} & \textbf{0.95} (22) & 0.50 (22) & 0.23 (22) \\
\hline
\end{tabular}
\caption{Accuracy results: predicting political slant as binary polarity, for all accounts available per method (`all'), or for the accounts represented by all methods (`common').}
\label{tab:polls_binary}
\end{table}

Table~\ref{tab:polls} reports the similarity of the poll-based rankings with the rankings generated using Eq.~\ref{eq:bias} and the different entity embedding schemes in terms of the Spearman's ranking correlation measure~\cite{hillCL15}. A perfect Spearman correlation of +1 indicates that the rankings are identical, where correlation of -1 would mean that the rankings are perfectly inverse. As shown in the table, the rankings produced using SocialVec are highly similar to the ground-truth rankings, as measured by the high correlation scores of 0.82 and 0.85 per the two polls. In contrast, the rankings produced by Wikipedia2Vec are not well-aligned with the ground-truth rankings, as indicated by the low Spearman correlation scores of 0.36 and 0.28. The rankings generated using the Wikidata graph embeddings are not meaningful altogether, yielding negative correlation scores.

Figure~\ref{fig:political} illustrates the distribution of the political orientation scores of the news sources included in the poll of 2020, as computed using SocialVec. The accounts are placed on the range of Democratic (left) to Republican (right), and are spaced along this range relative to their scores. Similar to the poll results, we observe that some news sources lie close to each other on this scale of political bias. Notably, gauging ranking similarity is highly sensitive, in that any differences in the ordering of accounts with similar scores are penalized. 

Table~\ref{tab:polls_binary} reports the results of a more lenient evaluation, where we consider the proportion of evaluated news accounts for which the polarity is correctly estimated. Here, we assign the computed political orientation to be conservative/Republican if the PO score is positive, and vice versa (Eq.~\ref{eq:bias}). As shown in the table, the binary political orientation predicted using SocialVec is accurate for 94\% and 97\% of the evaluated news accounts, where there exists a single mistake per the polls of 2014 and 2020, respectively. In contrast, Wikipedia2Vec embeddings yield low accuracy of 55\% and 60\% per those polls, and the Wikidata graph embeddings yield poor accuracies of 32\% and 27\%. To account for the differences in coverage, Table~\ref{tab:polls_binary} reports prediction accuracy also for the subset of accounts which are represented by all methods (`common'). As shown, the same trends persist. In error analysis, we found that faulty polarity predictions by SocialVec applied to news accounts for which the number of contexts (followers) in our sample of Twitter was the lowest among all the accounts included in each poll. If accounts with less than 5,000 followers in our data are removed from the evaluation, then SocialVec achieves perfect polarity classification results for all of the Twitter news accounts that are included in both of the reference polls. 

Overall, our results demonstrate that the political orientation of Twitter accounts can be accurately predicted based on the social contexts embedded in SocialVec. In contrast, the Wikipedia- and Wikidata-based entity embeddings fail to relate different entities by social aspects such as political affinity. We believe that the framework presented here for predicting political bias can be employed in future research for the assessment of various social biases using SocialVec.

\section{Conclusion}

We presented SocialVec, a framework for learning social entity embeddings from social networks, considering the accounts that users tend to co-follow as relevant contexts. We demonstrated the applicability of SocialVec embeddings in two case studies, obtaining competitive or SOTA results using minimal or no supervision, and showing advantageous performance over entity representations derived from knowledge bases.


There are naturally some limitations to our approach. An inherent limitation is that the social network of Twitter may provide a biased reflection of the real world. It has been shown, for example, that Twitter users are younger and more Democrat than the general public.\footnote{https://www.pewresearch.org/internet/2019/04/24/sizing-up-twitter-users}. In addition, while public figures like politicians and artists typically maintain popular Twitter accounts, some entity types, e.g., locations, may not be well-represented in Twitter, or invoke low interest in this platform. Furthermore, accounts may be banned from social networks like Twitter. Yet, we believe that SocialVec encapsulate valuable, wide and truthful social knowledge. Another inherent limitation of SocialVec, as any other embedding method, is that the quality of particular entity embeddings depends on sufficient context statistics for those entities. In the future, we plan to extend the scope of data that is modeled in SocialVec, where this would enable the learning of high-quality representations for entities that are popular locally, or within particular sub-communities.


This may be the first work to present social entity embeddings. We believe that the implications of this work go far and beyond the particular case studies presented in this work. We will make SocialVec embeddings publicly available, hoping to promote social knowledge modeling and exploration. 
As future research directions, we are interested to enhance the social knowledge encoded by SocialVec with account semantic types. We also wish to explore the integration of social context in content analysis, for example, for identifying entity mentions in tweets, and for the modeling of social context in applications of opinion mining.


\begin{small}
\bibliography{main}
\end{small}

\end{document}